\documentclass[%
 reprint,
superscriptaddress,
amsmath,amssymb,
aps,
floatfix,
]{revtex4-2}

\usepackage[utf8]{inputenc} 
\usepackage[T1]{fontenc}    
\usepackage{graphicx}       
\usepackage{dcolumn}        
\usepackage{bm}             
\usepackage{braket}         
\usepackage{booktabs}       
\usepackage{here}           
\usepackage[colorlinks=true, allcolors=blue]{hyperref} 
\usepackage{color}          
\usepackage{textcomp}       
\usepackage{comment}        
\usepackage{tabularx}       
\usepackage{mathptmx}
\usepackage{mathtools}
\usepackage{comment}

\newcommand{\nucleus}[2]{{}^{#1}\mathrm{#2}} 
\newcommand{\orbit}[1]{\mathrm{#1}} 

\begin{document}

\preprint{APS/123-QED}

\title{Systematic analysis of proton- and deuteron-induced one-proton knockout reactions}

\author{Hibiki Nakada}  %
\email[Email address: ]{nakada.hibiki@jaea.go.jp}
\affiliation{Research Center for Nuclear Physics, Osaka University, Ibaraki, Osaka 567-0047, Japan}
\affiliation{Department of Physics, Kyushu University, Fukuoka 819-0395, Japan}
\affiliation{Nuclear~Data~Center, Japan Atomic Energy Agency, Tokai, Ibaraki 319-1195, Japan}
\author{Shoya Ogawa}
\author{Yoshiki Chazono}
\author{Kazuyuki Ogata}  %
\affiliation{Department of Physics, Kyushu University, Fukuoka 819-0395, Japan}

             

\begin{abstract}
\noindent
\textbf{Background}:
The ratios of the one-proton knockout cross sections by a deuteron to those by a proton are about 1.5, indicating that using deuteron is more efficient than proton in yielding large knockout cross sections. However, this ratio differs from the intuitive expectation, and its underlying mechanism remains unclear.

\noindent
\textbf{Purpose}:
The purpose of this study is to clarify the mechanism behind the observed ratio by theoretically describing and analyzing the deuteron- and proton-induced one-proton knockout reactions.

\noindent
\textbf{Methods}:
Proton-induced one-proton knockout reactions are described within the standard distorted-wave impulse approximation (DWIA) framework, while deuteron-induced one-proton knockout reactions are treated with a new approach, DWIA-BU, that incorporates deuteron breakup into the DWIA.

\noindent
\textbf{Results}:
The ratios calculated with the DWIA-BU reproduce the experimental data reasonably, whereas those with the DWIA significantly underestimate them. The ratio of the corresponding elementary cross sections remains about 3.5 regardless of the energy, and the difference in absorption between the deuteron and the proton influences the ratios of knockout cross sections, resulting in agreement between the calculated ratios and the experimental data.

\noindent
\textbf{Conclusion}:
It is found that the deuteron breakup is essential to reproduce the experimental ratio. The ratios of the knockout cross sections are primarily determined by the difference in the elementary cross sections and that in the absorption between the deuteron and the proton.

\end{abstract}


\maketitle


\section{INTRODUCTION}\label{Sec_1}
The study of neutron-rich nuclei is essential for advancing our understanding of nuclear structure far from stability and has important implications for nuclear astrophysics. In particular, producing more neutron-rich nuclei is crucial for exploring the neutron drip line and probing nuclear matter under extreme isospin conditions; however, their short lifetimes make the production challenging. 
Thus far, nuclear fragmentation~\cite{Mocko2006,Alvarez2010} and fission~\cite{BERNAS1994,Armbruster2004} have served as the primary methods for generating neutron-rich nuclei~\cite{Thoennessen2004}. Neutron-rich nuclei produced via fragmentation have recently been used as beams in inverse kinematics, with proton knockout reactions~\cite{Thoennessen2003,Chant1983,Cowley1991,Kitching1967,Ogata2015,Minomo2017} expected to enable selective access to systems with even greater neutron excess.

To optimize the yield of neutron-rich nuclei produced in the proton knockout reaction, it is important to examine how the choice of reacting particle affects the knockout cross section. Quite recently, the ratios of the one-proton knockout cross sections by a deuteron to those by a proton for neutron-rich nuclei have been measured in inverse kinematics at around 240 MeV per nucleon~\cite{miwa2019}.
This measurement revealed that the ratios are approximately 1.5, indicating that using deuteron is more efficient than proton in view point the yield of neutron-rich nuclei. However, this enhancement is smaller than the naively expected factor of 4, based on the difference between the nucleon numbers of deuteron and proton, together with the identical property of two protons in the elementary process of the proton induced knockout reaction
 (see below). The origin of the ratios measured is thus not yet understood and requires theoretical investigation. 
In Ref.~\cite{miwa2019}, it is also shown that in two-proton knockout reactions, the deuteron-to-proton ratios are significantly large, although the mechanism is still not understood. While this open issue persists, the present study focuses on one-proton knockout reactions.

The purpose of this study is to clarify why the cross section for the one-proton knockout process by a deuteron, A($d,xp$)B, is about 1.5 times that of the process by a proton, A($p,2p$)B. Note that, for clarity, we treat proton and deuteron as a projectile in the symbolic notation, and we describe the reaction processes accordingly, i.e., assuming normal kinematics.  
We apply the distorted wave impulse approximation (DWIA) framework~\cite{wakasa2017,Ogata2024}, widely used for describing nucleon and particle knockout reactions.
In the A($d,xp$)B reaction, there are two components: 
A($d,dp$)B where the deuteron remains bound, and 
A($d,d^{\ast}p$)B where it breaks up. 
We describe the A($d,xp$)B reaction with an extended DWIA, DWIA-BU, which incorporates the deuteron breakup into the DWIA.

This paper is organized as follows. In Sec.~\ref{Sec_2}, we describe the A($p,2p$)B and A($d,dp$)B reactions with the DWIA, and introduce the DWIA-BU for the A($d,xp$)B reaction. In Sec.~\ref{Sec_3}, we compare the ratios of the calculated knockout cross sections for these reactions with experimental data and discuss the mechanism behind the ratio. Finally, a summary is given in Sec.~\ref{Sec_4}.


\section{FORMALISM}\label{Sec_2}
\subsection{DWIA framework}
\begin{figure}[h]
    \centering
    \includegraphics[width=0.8\hsize]{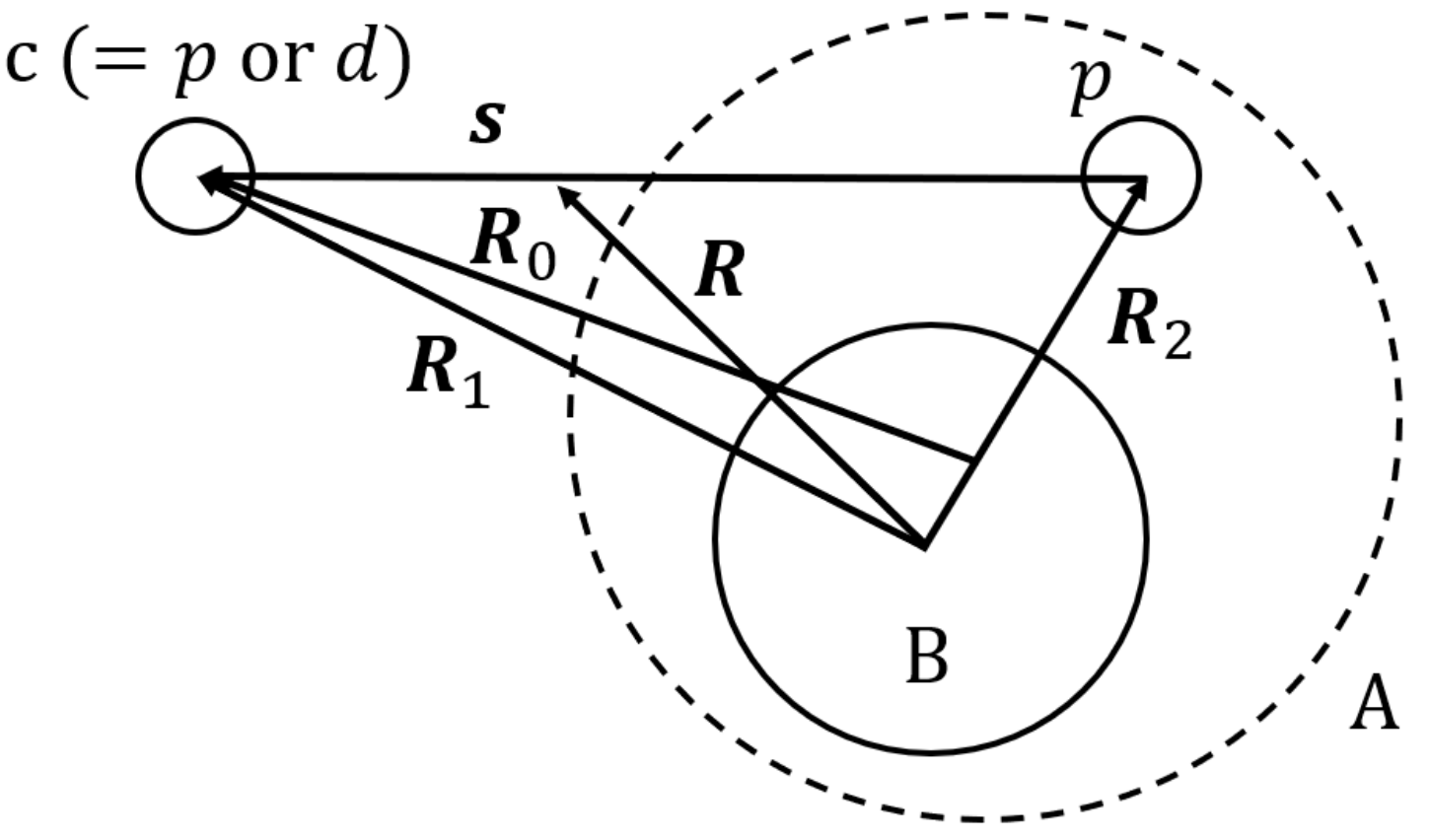}
    \caption{Definition of coordinates of the A($p,2p$)B and A($d,dp$)B reactions.}
\label{fig_coodir}
\end{figure}

We employ the DWIA framework~\cite{Chant1983,wakasa2017} to describe the A($p,2p$)B and A($d,dp$)B reactions. Figure~\ref{fig_coodir} shows the schematic illustration of the knockout reaction. The incoming (outgoing) proton or deuteron is denoted by particle $0~\text{(}1\text{)}$. The struck proton is labeled as particle 2, and the target (residual) nucleus is denoted by A (B).
In what follows, $\hbar \bm{K}_i$ and $E_i$ denote the momentum and total energy of particle $i$ ($= 0, 1, 2$), respectively.
The solid angle of the outgoing particle $j$ ($= 1~\text{or}~2$) is expressed as $\Omega_j$. Quantities with the superscript L are defined in the laboratory frame. While those without the superscript are defined in the center-of-mass frame.

In the DWIA framework, the transition matrices of the A($p,2p$)B and A($d,dp$)B reactions are expressed by
\begin{align}
    T_{\text{c}} &= \Braket{ \chi^{(-)}_{1,\bm{K}_1}(\bm{R}_1) \chi^{(-)}_{2,\bm{K}_2}(\bm{R}_2)  | t_{\text{c}p}(\bm{s}) | \varphi_p(\bm{R}_2) \chi^{(+)}_{0,\bm{K}_0}(\bm{R}_0) },
\label{eq_T_c}
\end{align}
where the subscript c specifies the incident particle ($p$ or $d$). The distorted wave for particle $i$ is represented by $\chi_i$. The superscripts $(+)$ and $(-)$ indicate that the distorted waves satisfy outgoing and incoming boundary conditions, respectively. $\varphi_p$ denotes the bound-state wave function of the struck proton, and $t_{\text{c}p}$ represents the effective interaction between particle ${\text{c}}$ and the proton.  

The coordinates $\bm{R}_1$, $\bm{R}_2$, and $\bm{R}_0$ can be rewritten in terms of $\bm{R}$ and $\bm{s}$, where $\bm{R}$ is the coordinate between B and the center of mass of particles 0 and 2, and $\bm{s}$ is the relative coordinate between particles 0 and 2:
\begin{align}
    \bm{R}_1 &= \bm{R} + \frac{1}{A_{\text{c}}+1}\bm{s}, \label{eq_R_1} \\
    \bm{R}_2 &= \bm{R} - \frac{A_{\text{c}}}{A_{\text{c}}+1}\bm{s}, \label{eq_R_2} \\
    \bm{R}_0 &= \bm{R} - \frac{1}{A}\bm{R} + \frac{A+A_{\text{c}}}{A} \frac{1}{A_{\text{c}}+1}\bm{s} \label{eq_R_0},
\end{align}
where $A_{\text{c}}$ and $A$ are the mass numbers of particle ${\text{c}}$ and $A$, respectively.
We substitute Eqs.~\eqref{eq_R_1}-\eqref{eq_R_0} into Eq.~\eqref{eq_T_c} and apply the asymptotic momentum approximation (AMA)~\cite{yoshida2016} to the distorted waves.  
In AMA, the propagation of the distorted wave over a short distance \(\Delta \bm{R}\) is approximated by a plane wave with the asymptotic momentum \(\hbar\bm{K}_i\):
\begin{align}
    \chi_{i,\bm{K}_i}(\bm{R} + \Delta \bm{R}) \approx \chi_{i,\bm{K}_i}(\bm{R}) e^{i \bm{K}_i \cdot \Delta \bm{R}}.
\end{align}
Accordingly, Eq.~\eqref{eq_T_c} can be rewritten as
\begin{align}
    T_{\text{c}} =& \tilde{t}_{\text{c}p}(\bm{\kappa}',\bm{\kappa}) \int d\bm{R}~\chi^{*(-)}_{1,\bm{K}_1}(\bm{R}) \chi^{*(-)}_{2,\bm{K}_2}(\bm{R}) \chi^{(+)}_{0,\bm{K}_0}(\bm{R})\\
    &\times e^{-i \bm{K}_0 \cdot \bm{R}/A} \varphi_p(\bm{R}),
\end{align}
where $\tilde{t}_{\text{c}p}(\bm{\kappa}',\bm{\kappa})$ is defined by
\begin{align}
    \tilde{t}_{\text{c}p}(\bm{\kappa}',\bm{\kappa}) \equiv \int d\bm{s}~e^{-i\bm{\kappa}'\cdot\bm{s}}  t_{\text{c}p}(\bm{s}) e^{i\bm{\kappa}\cdot\bm{s}} 
\label{eq_t_cp_tilde}
\end{align}
with the relative wavenumbers between particle ${\text{c}}$ and the proton in the initial and final state:
\begin{align}
    \bm{\kappa} &\equiv \frac{A+A_{\text{c}}}{A}\frac{1}{A_{\text{c}}+1}\bm{K}_0 - \frac{1}{A_{\text{c}}+1}\bm{K}_p,\\
    \bm{\kappa}' &\equiv \frac{1}{A_{\text{c}}+1}\bm{K}_1 - \frac{A_{\text{c}}}{A_{\text{c}}+1}\bm{K}_2.
\end{align}
Here, $\hbar\bm{K}_p$ is the momentum of the struck proton, which is determined to satisfy the total momentum conservation of the colliding particles.

Applying the on-the-energy-shell (on-shell) approximation to Eq.~\eqref{eq_t_cp_tilde}, we find
\begin{align}
    \left(\frac{\mu_{\text{c}p}}{2\pi\hbar^2}\right)^2|\tilde{t}_{\text{c}p}(\bm{\kappa}',\bm{\kappa})|^2\approx\frac{d\sigma_{\text{c}p}}{d\Omega_{\text{c}p}}(\theta^{\text{c}p},E^{\text{c}}),
\end{align}
where $\mu_{\text{c}p}$ denotes the reduced mass of the ${\text{c}}$-$p$ system.  $d\sigma_{\text{c}p}/d\Omega_{\text{c}p}$ refers to the elastic scattering cross section as a function of the scattering angle $\theta^{\text{c}p}$ and the incident energy $E^{\text{c}}$ of particle ${\text{c}}$ in the proton-rest frame.

With the AMA and on-shell approximation, the triple differential cross sections (TDXs) of the A($p,2p$)B and A($d,dp$)B reactions are given by
\begin{align}
\frac{d^3\sigma^{\text{DWIA}}_{\text{c}}}{dE_1^{\text{L}}\, d\Omega_1^{\text{L}}\, d\Omega_2^{\text{L}}}
    =& \frac{(2\pi)^4}{\hbar v}S_pF^{\text{L}}_{\text{kin}}\frac{E_1 E_2 E_{\text{B}}}{E^{\text{L}}_1 E^{\text{L}}_2 E^{\text{L}}_{\text{B}}}\frac{1}{2l+1}\nonumber\\
&\times\left(\frac{2\pi\hbar^2}{\mu_{\text{c}p}}\right)^2\frac{d\sigma_{\text{c}p}}{d\Omega_{\text{c}p}}(\theta^{\text{c}p}, E^{\text{c}})\, |\bar{T}_c|^2,
\label{eq_TDX}
\end{align}
\begin{align}
F^{\text{L}}_{\text{kin}} &\equiv \frac{E^{\text{L}}_1 K^{\text{L}}_1 E^{\text{L}}_2 K^{\text{L}}_2}{(\hbar c)^2}
\left[ 1 + \frac{E^{\text{L}}_2}{E^{\text{L}}_B} - \frac{E^{\text{L}}_2}{E^{\text{L}}_B} \frac{\bm{q}^{\text{L}} \cdot \bm{K}^{\text{L}}_2}{(K^{\text{L}}_2)^2} \right]^{-1}, \label{eq_Fkin}
\end{align}
and 
\begin{align}
    \bar{T}_{\text{c}} &= \int d\bm{R}~\chi^{*(-)}_{1,\bm{K}_1}(\bm{R}) \chi^{*(-)}_{2,\bm{K}_2}(\bm{R}) \chi^{(+)}_{0,\bm{K}_0}(\bm{R})  \varphi_p(\bm{R}),
\end{align}
where $E_{\text{B}}$ is the total energy of B, and $\bm{q}^{\text{L}}\equiv\bm{K}^{\text{L}}_{0}+\bm{K}^{\text{L}}_{\text{A}}-\bm{K}^{\text{L}}_{1}$ with the momentum of A in the laboratory frame, $\bm{K}^{\text{L}}_{\text{A}}$. The relative velocity between particle 0 and A is represented by $v$, the spectroscopic factor of the proton is denoted by $S_p$, and $l$ is the relative angular momentum between B and the proton in A.

In the plane wave impulse approximation (PWIA) that neglects the distorting potentials, $\bar{T}_c$ reduces to the Fourier transform of $\varphi_{p}(\bm{R})$:
\begin{align}
    \bar{T}_{\text{c}} = \int d\bm{R}~e^{i\bm{q}\cdot\bm{R}}  \varphi_p(\bm{R}),
\end{align}
where $\hbar \bm{q}$ is referred to as the missing momentum.

The total knockout cross section $\sigma_{\text{c}}$ is obtained by integrating the TDX (Eq.~\eqref{eq_TDX}) over $E^L_1$, $\Omega^L_1$, and $\Omega^L_2$:
\begin{align}
\sigma^{\text{DWIA}}_{\text{c}} = \int  \frac{d^3\sigma^{\text{DWIA}}_{\text{c}}}{dE_1^L d\Omega_1^L d\Omega_2^L} ~dE^L_1 d\Omega^L_1 d\Omega^L_2.
\label{eq_sigma_c}
\end{align} 

\subsection{DWIA incorporating deuteron breakup}
To describe the A($d,xp$)B reaction, we extend the DWIA to include the effect of the deuteron breakup.
Specifically, we add the breakup cross section to the $d$-$p$ elastic cross section as follows:
\begin{align}
\frac{d\sigma_{dp}}{d\Omega_{dp}} (\theta^{dp}, E^{d}) \rightarrow \frac{d\sigma_{dp}}{d\Omega_{dp}} (\theta^{dp}, E^{d})+\frac{\sigma^{\text{BU}}_{dp}}{4\pi}(E^{d}),
\label{eq_BU}
\end{align}
where the second term on right-hand side represents the breakup cross section of the $d$-$p$ reaction under the isotropic approximation. When only the first term on the right-hand side of Eq.~\eqref{eq_BU} is used as the elementary process in Eq.~\eqref{eq_TDX}, it corresponds to the A($d,dp$)B reaction, whereas using only the second term corresponds to the A($d,d^{\ast}p$)B reaction.
In what follows, the DWIA (PWIA) framework incorporating the deuteron breakup via Eq.~\eqref{eq_BU} is referred to as the DWIA-BU (PWIA-BU).
By applying Eq.~\eqref{eq_BU} to Eq.~\eqref{eq_TDX} and using Eq.~\eqref{eq_sigma_c}, we can obtain the knockout cross section of the A($d,xp$)B:
\begin{align}
    \sigma^{\text{DWIA-BU}}_d=\sigma^{\text{DWIA}}_{d}+\sigma^{\text{DWIA}}_{dd^{\ast}p},
\label{eq_sigma_DWIA-BU}
\end{align}
where the first and second terms on right-hand side are the knockout cross sections of the A($d,dp$)B and A($d,d^{\ast}p$)B, respectively.

We define the ratios of the knockout cross sections by a deuteron to those by a proton as:
\begin{align}
    \mathcal{R}^{\text{DWIA-BU}} \equiv \frac{\sigma^{\text{DWIA-BU}}_{d}}{\sigma^{\text{DWIA}}_p},\label{eq_ratio_DWIA-BU}\\
    \mathcal{R}^{\text{DWIA}} \equiv \frac{\sigma^{\text{DWIA}}_{d}}{\sigma^{\text{DWIA}}_p},\label{eq_ratio_DWIA}\\
    \mathcal{R}^{\text{PWIA-BU}} \equiv \frac{\sigma^{\text{PWIA-BU}}_{d}}{\sigma^{\text{PWIA}}_p}.\label{eq_ratio_PWIA-BU}
\end{align}

Note that, although the isotropic approximation in the second term on right-hand side of Eq.~\eqref{eq_BU} will be inadequate for TDX calculations, for which the kinematics of the elementary process are crucial,  
it is expected to provide reasonable results for the total knockout cross section obtained by integrating over a wide range of kinematics. The validity of the isotropic approximation is discussed in the Appendix~\ref{Appen_vali}.

\section{RESULTS AND DISCUSSION}\label{Sec_3}

\subsection{Numerical inputs}\label{subsec_nume}
We use a tabulated differential cross sections~\cite{chazono2022,yoshida2024} for the \(d\)-\(p\) elastic scattering,  
which was obtained by fitting experimental data~\cite{Ermisch2005,hatanaka2002,sekiguchi2002,sagara1994,Winkelmann1980,Boschitz1972,booth1971,Cahill1971,BUNKER1968,KURODA1964}  
of \(p\)-\(d\) scattering from 5 to 800 MeV, as functions of scattering energy and angle.  
We use the deuteron breakup cross section obtained by fitting the experimental data of the \(p\)-\(d\) total reaction cross section~\cite{Carlson1961,Catron1961,SEAGRAVE1972,brady1972} over the scattering energy range from 6 to 990 MeV. 
For the \(p\)-\(p\) scattering, the differential cross sections are obtained by assuming isotropic scattering for the total cross section given by an energy-dependent fit to experimental data~\cite{bertulani2010}.
This approach is adopted because the total knockout cross section is obtained by integrating the TDX, for which accurate $pp$ total cross sections are more important than a precise description of the differential cross section.

The proton distorted waves calculated with the EDAD1 parameter set of the Dirac phenomenology~\cite{hama1990,Cooper1993,Cooper2009}. The deuteron distorted waves are obtained with the folding potential, which is constructed by folding the proton- and neutron- distorting potentials with the density of the deuteron.
Nonlocality corrections to the distorted waves of the proton and deuteron are taken into account by using the Darwin factor~\cite{hama1990,Arnold1981} and the Perey factor~\cite{Perey1962}, respectively; the range of nonlocality for the deuteron is set to 0.54 fm~\cite{Igarashi1977}.

The bound-state wave function of the proton is obtained using the Woods-Saxon shaped potential by Bohr and Mottelson~\cite{Bohr1969}. The depth of the potential is adjusted to reproduce the proton separation energy. In the present calculation, as a first step, we assume that the residual nuclei are in the ground states. The single-particle orbit of the proton is chosen to conserve the spin-parites between the target nucleus and the system consisting of the residual nucleus and the proton, as summarized in Table~\ref{tab:nuclear_states}. When multiple orbits are possible, we adopt the orbit that is closer to the Fermi surface within the naive shell-model configuration.

In the present calculations, the DWIA calculations are performed using the computational code {\sc pikoe}~\cite{Ogata2024}.

\begin{table}[h!]
    \centering
    \caption{Spin-parites of the ground states of A and B taken from Refs.~\cite{Audi2017,Crawford2010,Ruiz2017,SUN2020}, and adopted orbits of the struck proton.}
    \renewcommand{\arraystretch}{1.2}
    \setlength{\tabcolsep}{6pt} 

    \begin{tabularx}{\linewidth}{>{\centering\arraybackslash}X >{\centering\arraybackslash}X >{\centering\arraybackslash}X}
        \toprule
        \textbf{A ($J^{\pi}$)} & \textbf{B ($J^{\pi}$)} & \textbf{Proton orbit} \\
        \midrule
            \multicolumn{1}{>{\raggedright\arraybackslash}p{2cm}}{\hspace*{0.4cm}\(\nucleus{51}{K} \;  (3/2^+)\)} & 
        \multicolumn{1}{>{\raggedright\arraybackslash}X}{\hspace*{0.4cm}\(\nucleus{50}{Ar} \; (0^+)\)} & 
        \multicolumn{1}{>{\raggedright\arraybackslash}X}{\hspace*{1cm}\(\orbit{0d_{3/2}}\)} \\
        
        \multicolumn{1}{>{\raggedright\arraybackslash}p{2cm}}{\hspace*{0.4cm}\(\nucleus{52}{Ca} \; (0^+)\)} & 
        \multicolumn{1}{>{\raggedright\arraybackslash}X}{\hspace*{0.4cm}\(\nucleus{51}{K} \; (3/2^+)\)} & 
        \multicolumn{1}{>{\raggedright\arraybackslash}X}{\hspace*{1cm}\(\orbit{0d_{3/2}}\)} \\
        
        \multicolumn{1}{>{\raggedright\arraybackslash}p{2cm}}{\hspace*{0.4cm}\(\nucleus{53}{Ca} \; (1/2^-)\)} & 
        \multicolumn{1}{>{\raggedright\arraybackslash}X}{\hspace*{0.4cm}\(\nucleus{52}{K} \; (2^-)\)} & 
        \multicolumn{1}{>{\raggedright\arraybackslash}X}{\hspace*{1cm}\(\orbit{0d_{3/2}}\)} \\
        
        \multicolumn{1}{>{\raggedright\arraybackslash}p{2cm}}{\hspace*{0.4cm}\(\nucleus{54}{Ca} \; (0^+)\)} & 
        \multicolumn{1}{>{\raggedright\arraybackslash}X}{\hspace*{0.4cm}\(\nucleus{53}{K} \; (3/2^+)\)} & 
        \multicolumn{1}{>{\raggedright\arraybackslash}X}{\hspace*{1cm}\(\orbit{0d_{3/2}}\)} \\
        
        \multicolumn{1}{>{\raggedright\arraybackslash}p{2cm}}{\hspace*{0.4cm}\(\nucleus{55}{Sc} \; (7/2^-)\)} & 
        \multicolumn{1}{>{\raggedright\arraybackslash}X}{\hspace*{0.4cm}\(\nucleus{54}{Ca} \; (0^+)\)} & 
        \multicolumn{1}{>{\raggedright\arraybackslash}X}{\hspace*{1cm}\(\orbit{0f_{7/2}}\)} \\
        
        \multicolumn{1}{>{\raggedright\arraybackslash}p{2cm}}{\hspace*{0.4cm}\(\nucleus{56}{Sc} \; (1^+)\)} & 
        \multicolumn{1}{>{\raggedright\arraybackslash}X}{\hspace*{0.4cm}\(\nucleus{55}{Ca} \; (5/2^-)\)} & 
        \multicolumn{1}{>{\raggedright\arraybackslash}X}{\hspace*{1cm}\(\orbit{0f_{7/2}}\)} \\
        
        \multicolumn{1}{>{\raggedright\arraybackslash}p{2cm}}{\hspace*{0.4cm}\(\nucleus{57}{Sc} \; (7/2^-)\)} & 
        \multicolumn{1}{>{\raggedright\arraybackslash}X}{\hspace*{0.4cm}\(\nucleus{56}{Ca} \; (0^+)\)} & 
        \multicolumn{1}{>{\raggedright\arraybackslash}X}{\hspace*{1cm}\(\orbit{0f_{7/2}}\)} \\
        
        \multicolumn{1}{>{\raggedright\arraybackslash}p{2cm}}{\hspace*{0.4cm}\(\nucleus{58}{Ti} \; (0^+)\)} & 
        \multicolumn{1}{>{\raggedright\arraybackslash}X}{\hspace*{0.4cm}\(\nucleus{57}{Sc} \; (7/2^-)\)} & 
        \multicolumn{1}{>{\raggedright\arraybackslash}X}{\hspace*{1cm}\(\orbit{0f_{7/2}}\)} \\
        
        \multicolumn{1}{>{\raggedright\arraybackslash}p{2cm}}{\hspace*{0.4cm}\(\nucleus{59}{Ti} \; (1/2^-)\)} & 
        \multicolumn{1}{>{\raggedright\arraybackslash}X}{\hspace*{0.4cm}\(\nucleus{58}{Sc} \; (3^+)\)} & 
        \multicolumn{1}{>{\raggedright\arraybackslash}X}{\hspace*{1cm}\(\orbit{0f_{7/2}}\)} \\
        
        \multicolumn{1}{>{\raggedright\arraybackslash}p{2cm}}{\hspace*{0.4cm}\(\nucleus{60}{Ti} \; (0^+)\)} & 
        \multicolumn{1}{>{\raggedright\arraybackslash}X}{\hspace*{0.4cm}\(\nucleus{59}{Sc} \; (7/2^-)\)} & 
        \multicolumn{1}{>{\raggedright\arraybackslash}X}{\hspace*{1cm}\(\orbit{0f_{7/2}}\)} \\
        
        \multicolumn{1}{>{\raggedright\arraybackslash}p{2cm}}{\hspace*{0.4cm}\(\nucleus{61}{V} \;  (3/2^-)\)} & 
        \multicolumn{1}{>{\raggedright\arraybackslash}X}{\hspace*{0.4cm}\(\nucleus{60}{Ti} \; (0^+)\)} & 
        \multicolumn{1}{>{\raggedright\arraybackslash}X}{\hspace*{1cm}\(\orbit{1p_{3/2}}\)} \\
        
        \multicolumn{1}{>{\raggedright\arraybackslash}p{2cm}}{\hspace*{0.4cm}\(\nucleus{62}{V} \;  (3^+)\)} & 
        \multicolumn{1}{>{\raggedright\arraybackslash}X}{\hspace*{0.4cm}\(\nucleus{61}{Ti} \; (1/2^-)\)} & 
        \multicolumn{1}{>{\raggedright\arraybackslash}X}{\hspace*{1cm}\(\orbit{0f_{7/2}}\)} \\
        \bottomrule
    \end{tabularx}
    \label{tab:nuclear_states}
\end{table}

\subsection{Comparison with calculation results and experimental data}
\begin{figure}[h]
    \centering
    \includegraphics[width=0.98\hsize]{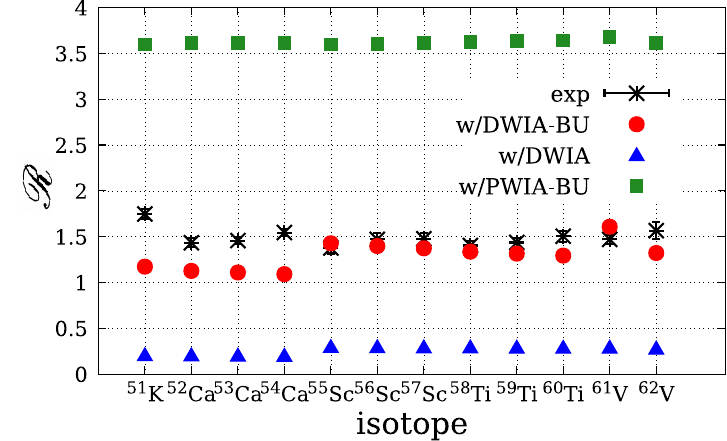}
    \caption{The ratios of the cross sections induced by a deuteron to those by a proton for the 12 neutron-rich nuclei at around 240 MeV per nucleon. The filled circles, the filled triangles, and the filled square represent the results obtained with the DWIA-BU, DWIA, and PWIA-BU. Experimental data are taken from Ref.~\cite{miwa2019}.}
\label{fig_ratio_tag-ave}
\end{figure}

We compare  the numerical results of $\mathcal{R}$ with the experimental data~\cite{miwa2019} for the neutron-rich nuclei considered in Fig.~\ref{fig_ratio_tag-ave}. Note that $\mathcal{R}$ is independent of the spectroscopic factor.
The asterisk indicates the experimental data, with an initial incident energy of 240~MeV per nucleon; the cross sections are averaged over the incident energy ranging from 150 to 240~MeV due to beam energy loss in the target. The filled circles, triangles, and squares represent the results calculated with the DWIA-BU, DWIA, and PWIA-BU, respectively. The energy averaging is performed in the same way as for the experimental data.

$\mathcal{R}$ calculated with the DWIA-BU reproduces the experimental data reasonably well. Slight undershooting is observed for $^{51}\mathrm{K}$–$^{54}\mathrm{Ca}$. For $^{61}\mathrm{V}$, the DWIA-BU result is slightly larger than for the neighboring nuclei.
As seen in Table~\ref{tab:nuclear_states}, this difference may be attributed to the difference in  the orbital angular momentum of the struck proton, as discussed in detail in Sec.~\ref{subsec_orbit}.

In contrast, the results calculated with the DWIA significantly underestimate the experimental data.
By comparing the DWIA-BU and DWIA results, one sees clearly that the inclusion of the breakup component of deuteron is essential for reproducing the experimental data. The results with the PWIA-BU are about 3.5 regardless of the nuclei; this value is explained by the difference between the elementary cross sections for the A$(d,xp)$B and A$(p,2p)$B reactions (see Sec.~\ref{subsec_rela}).

From the comparison of the PWIA-BU and DWIA-BU results, it is clear that the nuclear absorption reduces the ratios. This is because a deuteron is more strongly affected by the absorption than a proton.

To match the experimental condition, the theoretical results shown in Fig.~\ref{fig_ratio_tag-ave} are averaged over the incident energy from 150 to 240 MeV per nucleon. However, it is found that the results are essentially the same as those at 240 MeV per nucleon. Therefore, in the following discussion in Secs.~\ref{subsec_rela} and \ref{subsec_orbit}, we focus on the results at 240 MeV per nucleon.
\subsection{Relation with the ratio and elementary process}
\label{subsec_rela}


Here, we discuss the mechanism underlying the values of $\mathcal{R}$ with the PWIA-BU ($\approx$3.5) and DWIA-BU ($\approx$1.4) as shown in Fig~\ref{fig_ratio_tag-ave}. For simplicity, in the following analysis we make the isotropic approximation
to the $dp$ and $pN$ elementary processes. 
Then, we express Eq.\eqref{eq_sigma_c} in a
discrete form:%
\begin{equation}
\sigma_{\mathrm{c}}\approx\bar{\sigma}_{\mathrm{c}}\rightarrow\sum_{i}%
\bar{\sigma}_{i}^{\mathrm{c}},%
\label{eq_sigma_i}
\end{equation}
with%
\begin{equation}
\bar{\sigma}_{i}^{\mathrm{c}}=\left(  \frac{d^{3}\bar{\sigma}_{\mathrm{c}}%
}{dE_{1i}^{\mathrm{L}}d\Omega_{1i}^{\mathrm{L}}d\Omega_{2i}^{\mathrm{L}}%
}\right)  \Delta E_{1}^{\mathrm{L}}\sin\theta_{1i}^{\mathrm{L}}\Delta
\theta_{1}^{\mathrm{L}}\Delta\phi_{1}^{\mathrm{L}}\sin\theta_{2i}^{\mathrm{L}%
}\Delta\theta_{2}^{\mathrm{L}}\Delta\phi_{2}^{\mathrm{L}}\label{sigibar1}%
\end{equation}
and the bar over $\sigma$ indicates that it is obtained with the isotropic
approximation. One may rewrite $\bar{\sigma}_{i}$ as%
\begin{equation}
\bar{\sigma}_{i}^{\mathrm{c}}=\frac{\sigma_{\mathrm{c}p}^{\mathrm{tot}}\left(
E_{i}^{\mathrm{c}}\right)  }{4\pi}w_{i}^{\mathrm{c}}.\label{sigibar2}%
\end{equation}
The weight factor $w_{i}^{\mathrm{c}}$ is defined by Eqs. (\ref{sigibar1}) and
(\ref{sigibar2}). The index $i$ represents a set of the five kinematical
variables, which uniquely determines $E_{i}^{\mathrm{c}}$. Note that the
discrete form is used in the actual numerical calculations.

Here, we rewrite $\bar{\sigma}_{\mathrm{c}}$ with sorting $\bar{\sigma}%
_{i}^{\mathrm{c}}$ by $E_{i}^{\mathrm{c}}$ as follows:%
\begin{align}
\bar{\sigma}_{\mathrm{c}}  &  =\sum_{j}\sum_{\substack{i\\E_{i}^{\mathrm{c}%
}\in\left[  \varepsilon_{j}-\Delta\varepsilon,\varepsilon_{j}+\Delta
\varepsilon\right]  }}\frac{\sigma_{\mathrm{c}p}^{\mathrm{tot}}\left(
E_{i}^{\mathrm{c}}\right)  }{4\pi}w_{i}^{\mathrm{c}}\nonumber\\
&  \approx\sum_{j}\frac{\sigma_{\mathrm{c}p}^{\mathrm{tot}}\left(
\varepsilon_{j}\right)  }{4\pi}\sum_{\substack{i\\E_{i}^{\mathrm{c}}%
\in\left[  \varepsilon_{j}-\Delta\varepsilon,\varepsilon_{j}+\Delta
\varepsilon\right]  }}w_{i}^{\mathrm{c}}\equiv\sum_{j}\frac{\sigma
_{\mathrm{c}p}^{\mathrm{tot}}\left(  \varepsilon_{j}\right)  }{4\pi}%
W_{j}^{\mathrm{c}}.
\label{eq_weight}
\end{align}
It has been assumed that the $E_{i}^{\mathrm{c}}$ dependence of
$\sigma_{\mathrm{c}p}^{\mathrm{tot}}$ within $\left[  \varepsilon_{j}%
-\Delta\varepsilon,\varepsilon_{j}+\Delta\varepsilon\right]  $ is weak. 
Similarly, within DWIA-BU, the total knockout cross section can be expanded as in Eqs.~\eqref{eq_sigma_i}-\eqref{eq_weight}.
In the following analysis, we set $\varepsilon_{j}=10j$ MeV for $j=1$--$49$ and
$\Delta\varepsilon=10$ MeV.

\begin{figure}[h]
    \centering
    \includegraphics[width=1.05\hsize]{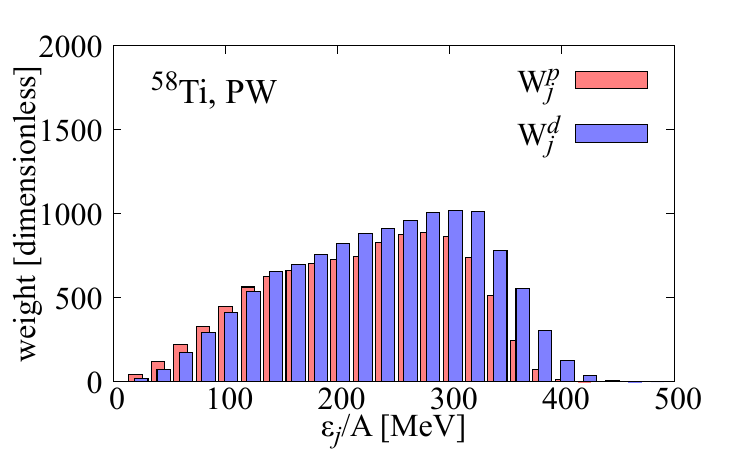}
    \caption{Energy distributions of the weights $W^{p}_{j}$ (red) and $W^{d}_{j}$ (blue) for $^{58}\mathrm{Ti}$, calculated with the PWIA and PWIA-BU, respectively. The weights $W^{p}_{j}$ and $W^{d}_{j}$ are defined in Eq.~\eqref{eq_weight}. The horizontal axis shows the energy per nucleon of the elementary process in the proton-rest frame}
\label{Fig_hist_58Ti_PW}
\end{figure}

\begin{figure}[h]
    \centering
    \includegraphics[width=0.9\hsize]{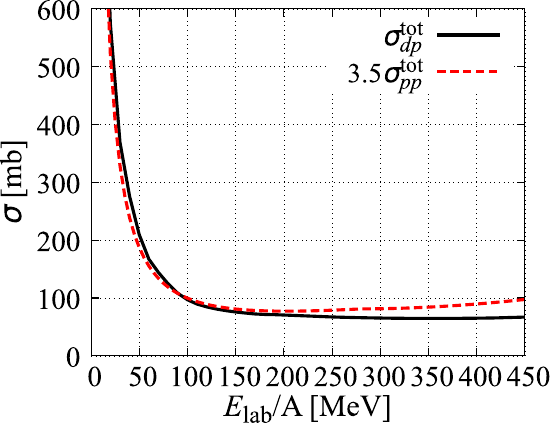}
    \caption{$dp$ total cross section $\sigma_{dp}^{\rm tot}$ (black solid line) and $pN$ total cross section $\sigma_{pp}^{\rm tot}$ (red dashed line) as a function of the incident energy per nucleon in the proton-rest frame. As for $\sigma_{pp}^{\rm tot}$, the parametrization in Ref.~\cite{bertulani2010} is adopted.
}
\label{Fig_elem_C3_5}
\end{figure}

Figure~\ref{Fig_hist_58Ti_PW} shows $W^{p}_{j}$ and $W^{d}_{j}$ on $ ^{58}\mathrm{Ti}$ calculated with the PWIA and PWIA-BU, respectively. 
One can see that $ W^{p}_{j} $ and $ W^{d}_{j} $ are almost identical in both magnitude and shape across the entire energy range. Thus, $\mathcal{R}$ with PWIA-BU is governed by the difference in the elementary cross sections in the PW limit. We then compare $ \sigma^{\text{tot}}_{dp} $ and $ \sigma^{\text{tot}}_{pp} $ in Fig.~\ref{Fig_elem_C3_5}; clearly, $ \sigma^{\text{tot}}_{dp} \approx 3.5 \sigma^{\text{tot}}_{pp} $ holds in the whole energy region considered. 
The value can be derived using the optical theorem and effective $N$-$N$ interactions (see Appendix~\ref{Appen_theorem}).
These results suggest that $\mathcal{R}$ with the PWIA-BU calculation reflects the ratio of $\sigma^{\text{tot}}_{dp}$ to $\sigma^{\text{tot}}_{pp}$.

\begin{figure}[h!]
    \centering
    \includegraphics[width=1.05\hsize]{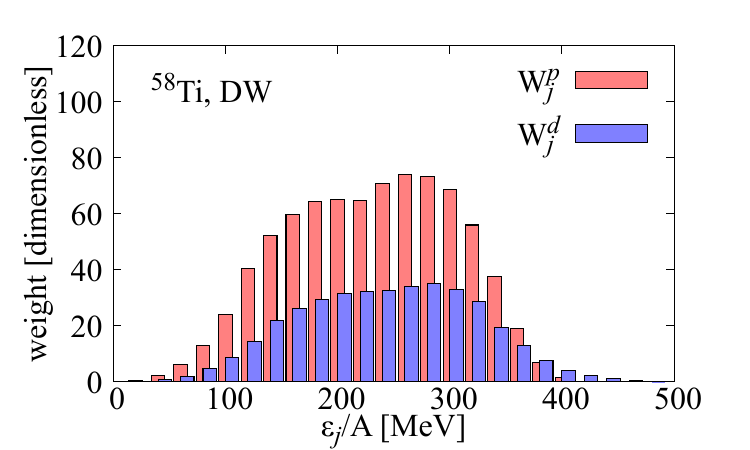}
    \caption{Same as Fig.~\ref{Fig_hist_58Ti_PW} but calculated with the DWIA and DWIA-BU.}
\label{Fig_hist_58Ti_DW}
\end{figure}

In the cases of the DWIA and DWIA-BU, as shown in Fig.~\ref{Fig_hist_58Ti_DW}, $W^{d}_{j}$ is smaller than $W^{p}_{j}$ because of the stronger absorption than for the proton. It is found that $W^{d}_{j}/W^{p}_{j} \approx 0.4$ in the region where $W_j^p$ and $W_j^d$ are mainly distributed. By considering the ratio of the weights and that of the elementary cross sections, $\sigma^{\text{tot}}_{dp}/\sigma^{\text{tot}}_{pp} \approx 3.5$ together, $\mathcal{R}$ calculated with the DWIA-BU is estimated to be $0.4 \times 3.5=1.4$. Thus, the essential feature of the experimental data is explained well.

\subsection{Dependence of the ratio calculated with DWIA-BU}\label{subsec_orbit}
\begin{figure}[h]
    \centering
    \includegraphics[width=1\hsize]{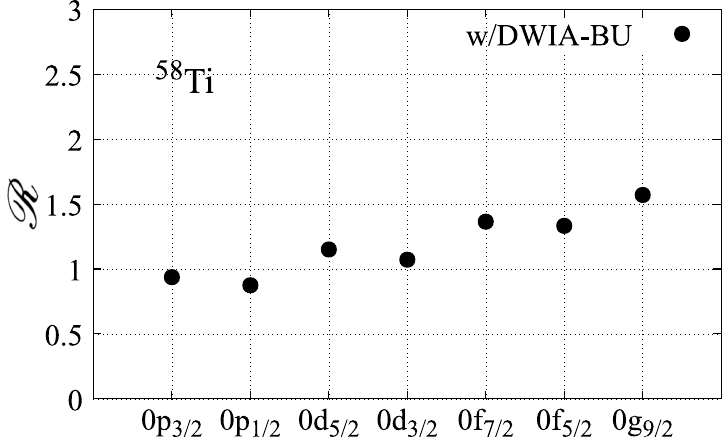}
    \caption{$l$-dependence of $\mathcal{R}$ calculated with the DWIA-BU for $ {^{58}\text{Ti}} $.}
\label{Fig_ratio_58Ti}
\end{figure}

\begin{figure}[h]
    \centering
    \includegraphics[width=1\hsize]{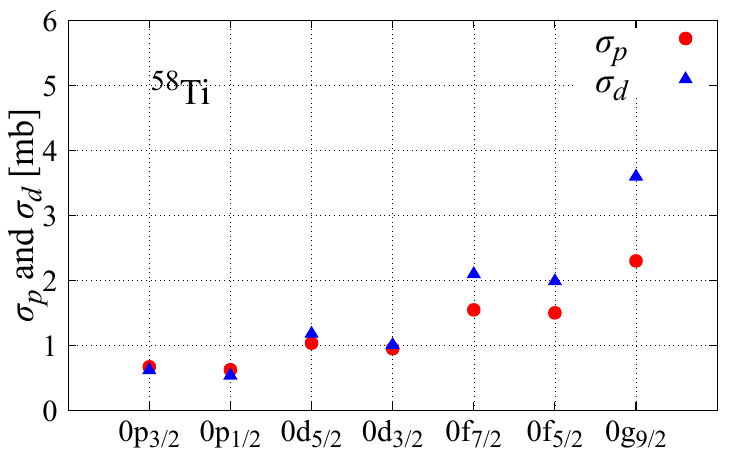}
    \caption{The one-proton knockout cross sections, $\sigma_{p}$ and $\sigma_{d}$, calculated with the DWIA and DWIA-BU, respectively, for different orbits in $^{58}\text{Ti}$. The spectroscopic factor is set to unity. The horizontal axis represents the orbit of the struck proton.}
\label{Fig_sigma_58Ti}
\end{figure}

\begin{figure}[h]
    \centering
    \includegraphics[width=1\hsize]{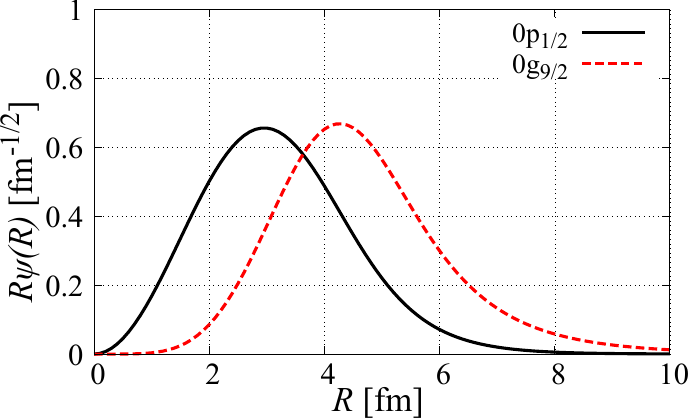}
    \caption{Radial distributions of the bound-state wave functions $ R\psi(R) $ for the $ 0\text{p}_{1/2} $ orbit shown as the solid line and the $ 0\text{g}_{9/2} $ orbit shown as the dashed line.}
\label{Fig_wf_58Ti_R2}
\end{figure}

As mentioned in Sec.~\ref{Sec_2}, the difference in $\mathcal{R}$ calculated with the DWIA-BU among the target nuclei may be attributed to the orbital angular momentum $l$ of the struck proton. 
To investigate the $l$-dependence more clearly, we show in Fig.~\ref{Fig_ratio_58Ti} $\mathcal{R}$ calculated with the DWIA-BU for $^{58}\text{Ti}$, with varying $l$.
One sees that $\mathcal{R}$ increases as $l$ becomes larger. In Fig.~\ref{Fig_sigma_58Ti}, we show the $l$-dependence of $\sigma_{p}$ and $\sigma_{d}$ individually. It is found that both cross sections increase as $l$ increases, but $\sigma_d$ has a stronger $l$-dependence.
This trend can be attributed to the radial distribution of the bound-state wave functions $\varphi_p$. Figure~\ref{Fig_wf_58Ti_R2} shows $\varphi_p$ in $^{58}\text{Ti}$, assuming the $0\text{p}_{1/2}$ (solid line) and $0\text{g}_{9/2}$ (dashed line) orbits. 
By comparing the two, one finds that the wave function for $0\text{g}_{9/2}$ is distributed more outward than that for $0\text{p}_{1/2}$. 
The outward shift makes the proton knockout process less affected by the nuclear absorption. In other words, as $l$ decreases, the effect of the absorption becomes stronger, leading to a more pronounced reduction in $\sigma_d$ compared to $\sigma_p$. Consequently, $\mathcal{R}$ increases as $l$ becomes larger, which matches the trend of $\mathcal{R}$ shown by the filled circles in Fig.~\ref{fig_ratio_tag-ave}.

\section{SUMMARY}\label{Sec_4}
We have analyzed the proton- and deuteron-induced one-proton knockout reactions, i.e., A($p,2p$)B and A($d,xp$)B.
To take into account the effect of deuteron breakup, we describe the A($d,xp$)B reaction with the DWIA-BU (PWIA-BU), which is the DWIA (PWIA) framework incorporating the deuteron breakup on the elementary $d$-$p$ process.
We have calculated the ratios of the knockout cross sections induced by a deuteron to those by a proton on several neutron-rich nuclei. The ratios calculated with the DWIA-BU are about 1.4, which reasonably reproduce the experimental data. The ratios with the DWIA are approximately 0.2, which significantly underestimate the experimental data. From these results, it is found that including the breakup effect on the elementary process is crucial for reproducing the experimental data.

In contrast, the ratios calculated with the PWIA-BU, which neglect nuclear absorption, are approximately 3.5 and independent of the target nucleus. This value is understood to be the ratio $\sigma^{\text{tot}}_{dp}/\sigma^{\text{tot}}_{pp}$. A detailed analysis clarified that the reduction of the ratio from 3.5 to about 1.4 is due to the stronger absorption of the deuteron than that of the proton.

We also investigated the dependence of the ratio calculated with DWIA-BU on the orbital angular momentum $l$ of the struck proton. For $^{58}\mathrm{Ti}$, the ratio increases with $l$. This is because the bound-state wave function shifts outward for higher $l$, suppressing absorption for both the proton and the deuteron, but especially for the latter. As a result, $\sigma_d$ is more sensitive to $l$ than $\sigma_p$, leading to an increase of the ratio with $l$.

For more precise comparisons with experimental data, it is necessary to take into account all possible configurations of the orbits of struck proton and the residual nuclear states.
On another front, the present DWIA-BU is not limited to deuteron-induced reactions, but can be extended to inclusive reactions involving other projectiles.

\appendix
\section{Validity of isotropic assumption for the $d$-$p$ breakup cross section}\label{Appen_vali}

\begin{figure}[htbp]
    \centering
    \includegraphics[width=0.95\hsize]{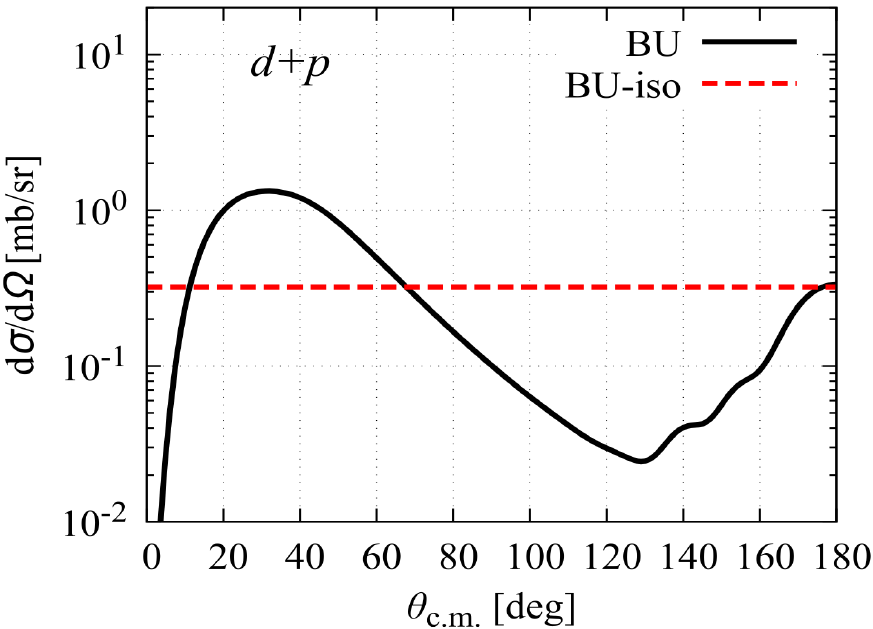}
    \caption{Angular distribution of the $d$-$p$ breakup cross section (black solid line) taken from~\cite{chazono2022} at 250 MeV per nucleon and its angular average (red dashed line).}
    \label{fig_BUiso_CS}
\end{figure}

We examine the validity of the isotropic approximation for the $d$-$p$ breakup cross section employed in the DWIA-BU. In Fig.~\ref{fig_BUiso_CS}, we compare the deuteron breakup cross section to the $pn$ s-wave breakup states (black solid line) at 250 MeV per nucleon, which is taken from Ref.~\cite{chazono2022}, and its angular average (red dashed line).
\begin{figure}[htbp]
    \centering
    \includegraphics[width=1.0\hsize]{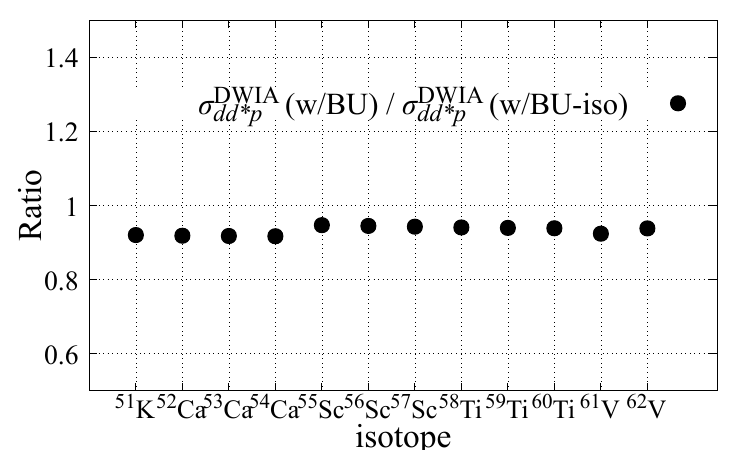}
    \caption{The ratios of the deuteron-induced one-proton knockout cross sections at 250 MeV per nucleon using the $d$-$p$ breakup cross section with angular dependence and its isotropic one.}
    \label{fig_BUiso_Ratio}
\end{figure}
Figure~\ref{fig_BUiso_Ratio} shows the ratio of $\sigma^{\text{DWIA}}_{dd*p}$ at 250 MeV calculated using $dp$ breakup cross section with the angular dependence to that using the isotropic one. In this calculation, the variation of the incident energy of the elementary process is disregarded for simplicity.
From this result, the ratios are about 0.93, indicating that the effect of the isotropic approximation is only a few percent.

\section{Derivation of the ratio of $\sigma^{\text{tot}}_{dp}$ to $\sigma^{\text{tot}}_{pp}$}\label{Appen_theorem}

By using the optical theorem and effective $N$-$N$ interactions, we derive the ratio of $\sigma^{\text{tot}}_{dp}$ to $\sigma^{\text{tot}}_{pp}$. 
In this derivation, the bound-state wave function of the deuteron is assumed to be a delta function; however, assuming a Gaussian function instead yields the same result. For simplicity, the Coulomb interaction is neglected.

By using the effective interactions $t_{pp}$ and $t_{np}$, the $pp$ and $dp$ scattering cross sections can be written as follows:
\begin{align}
\frac{d\sigma_{pp}}{d\Omega_{pp}} \approx \frac{\mu_{pp}^2}{(2\pi \hbar^2)^2} 
\left| \langle e^{i\bm{\kappa}'\cdot\bm{R}} | t_{pp} | e^{i\bm{\kappa}\cdot\bm{R}} \rangle \right|^2, \label{eq_1}\\[10pt]
\frac{d\sigma_{dp}}{d\Omega_{dp}} \approx \frac{\mu_{dp}^2}{(2\pi \hbar^2)^2} 
\left| \langle e^{i\bm{\kappa}'\cdot\bm{R}} 
\Phi(\bm{r}) | t_{pp} + t_{np} | \Phi(\bm{r}) e^{i\bm{\kappa}\cdot\bm{R}} \rangle \right|^2,\label{eq_2}
\end{align}
where $\bm{R}$ denotes the coordinate of the proton (or deuteron) relative to the nucleon, and $\bm{r}$ denotes the relative coordinate between the two nucleons within the deuteron. $\Phi(\bm{r})$ is the bound-state wavefunction of the deuteron. $\kappa$ and $\kappa'$ are the relative momenta between the proton (or deuteron) and the nucleon in the initial and final states, respectively. 

The scattering amplitude for $pp$ scattering is expressed as
\begin{align}
f_{pp}(\theta) &= -\frac{\mu_{pp}}{2\pi \hbar^2} 
\int d\bm{R} \, e^{-i\bm{\kappa}'\cdot\bm{R}} e^{i\bm{\kappa}\cdot\bm{R}} t_{pp}(R),
\end{align}
and for $\bm{\kappa}' = \bm{\kappa}$, one obtains
\begin{align}
f_{pp}(0) &= -\frac{\mu_{pp}}{2\pi \hbar^2} 
\int d\bm{R} \, t_{pp}(R).
\label{eq_f0pp}
\end{align}
Similarly, the scattering amplitude for $dp$ scattering is written as
\begin{align}
f_{dp}(\theta) &= -\frac{\mu_{dp}}{2\pi \hbar^2} 
\int d\bm{R} \, e^{-i\bm{\kappa}'\cdot\bm{R}} e^{i\bm{\kappa}\cdot\bm{R}} 
\int d\bm{r} \, |\Phi(\bm{r})|^2 (t_{pp} + t_{np}),
\end{align}
and for $\bm{\kappa}'=\bm{\kappa}$,
\begin{align}
f_{dp}(0) &= -\frac{\mu_{dp}}{2\pi \hbar^2} 2 \int d\bm{R} \, t_{pp}(R),
\label{eq_f0dp}
\end{align}
where $|\Phi(\bm{r})|^2 \to \delta(\bm{r})$ and $t_{np} \approx t_{pp} $ is adopted.

The optical theorem is generally expressed as $\text{Im}f(0)=(C_{\text{id}}k /4\pi )\sigma^{\text{tot}}$. The factor $C_{\text{id}}$ accounts for whether the scattering particles are identical or non-identical; $C_{\text{id}}=2~(1)$ for identical (non-identical) particles.
Finally, applying this to Eqs.~\eqref{eq_f0pp} and \eqref{eq_f0dp} and taking their ratio, one can obtain
\begin{align}
    \frac{\sigma^{\text{tot}}_{dp}}{\sigma^{\text{tot}}_{pp}}&=\frac{C^{pp}_{\text{id}}}{C^{dp}_{\text{id}}}\frac{k_p}{k_d}\frac{\text{Im}[f_{dp}(0)]}{\text{Im}[f_{pp}(0)]}\approx 4,
\end{align}
where \( m_d \approx 2 m_p \) and the incident energies per nucleon of the proton and deuteron are assumed to be the same.

\begin{acknowledgments}
This work was supported by JST, the establishment of university fellowships towards the creation of science technology innovation (Grant No. JPMJFS2125), and by Grants-in-Aid of the Japan Society for the Promotion of Science (Grants Nos. JP21H04975 and JPMJER2304).
The computation was carried out with the computer facilities at the Research Center
for Nuclear Physics, Osaka University.
\end{acknowledgments}


\bibliographystyle{apsrev4-2}
\bibliography{prc_knock}

\end{document}